# V2X Misbehavior and Collective Perception Service: Considerations for Standardization

Mohammad Raashid Ansari, Jean-Philippe Monteuuis, Jonathan Petit, Cong Chen

Qualcomm Technologies, Inc.

Boxborough, MA, USA

{ransari,jmonteuu,petit,congchen}@qti.qualcomm.com

*Abstract*—Connected and Automated Vehicles (CAV) use sensors and wireless communication to improve road safety and efficiency. However, attackers may target Vehicle-to-Everything (V2X) communication. Indeed, an attacker may send authenticated-but-wrong data to send false location information, alert incorrect events or report a bogus object endangering other CAVs' safety. Currently, Standardization Development Organizations are working on developing security standards against such attacks. Unfortunately, current standardization efforts do not include misbehavior specifications for advanced V2X services such as Collective Perception (CP) yet. This work assesses the security of CP Messages and proposes inputs for consideration in existing standards.

## I. INTRODUCTION

Vehicle-to-Everything (V2X) communication has the potential to tremendously improve vehicle safety technology to its next evolution. It enables V2X equipped vehicles to exchange their telematics information to create awareness, especially in non-line-of-sight (NLoS) conditions. This is achieved by broadcasting a message called Basic Safety Message (BSM), or Cooperative Awareness Message (CAM). Both of these messages contain the same information (location and kinematic state of the sender) but are defined in two different standards. BSM is defined in the Society of Automotive Engineers (SAE) J2735 standard [1] and CAM is defined in the European Telecommunications Standards Institute (ETSI) European Standard (EN) 302 637-2 standard [2]. However, not all objects on the road may be equipped with V2X capability (e.g., non-V2X vehicles, pedestrians, obstacles, animals). Therefore, a new service called Collective Perception Service (CPS) has been created to allow sharing observations of such objects by sharing sensor data among V2X-enabled vehicles about non-V2X objects on the road. Vehicles participating to the CPS generate and consume a message called Collective Perception Message (CPM) and is designed to complement the BSM/CAM service. Henceforth, we will refer to both BSM and CAM services as only CAM service.

CAMs and CPMs are supposed to be used to make driving decisions by an operator and/or an automated driving system. Due to this reason, these services become safety-critical and it is paramount that the information passed through these services is accurate. An attacker similar to the one discussed in Section IV can send wrong data in order to affect receivers' telematics awareness negatively. It has been shown that attacks on BSMs can have dramatic effects on V2X applications [3]. To detect and protect against such attackers, a Misbehavior Detection System (MBDS) must be deployed [4]. However, very little research has been done on the security of CPMs. In this paper, we summarize the results of a threat assessment (TA) on the Collective Perception Service (CPS) defined in ETSI Technical Report (TR) 103 562 [5] and ETSI Technical Standard (TS) 103 324 [6]. We also identify gaps in the current standardization of CPS and MBDS and propose items for consideration.

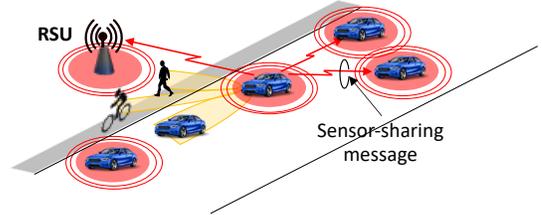

Fig. 1: Collective Perception Scenario

The paper is organized as follows. Section II presents the standardization and academic efforts in the domain of V2X CP and its security. Section III details the system model and the CPM. Section IV describes the attacker model considered in our TA presented in Section V. Section VI discusses the open challenges facing standardization and research to achieve a secure CPS. Finally, Section VII concludes this paper.

## II. RELATED WORK

This section provides an overview of functional and security standards for CPM. Additionally, this section includes related academic work.

### A. Standardization

In this section, we briefly introduce existing and ongoing standards from a functional and security perspective.

*1) Functional Standards:* The notion of CP has been introduced in the V2X community to share perceived scene information among V2X agents and smart infrastructures. As a result, each V2X agent can enhance its detection capabilities by considering the received information over the V2X network about non-V2X objects, which are beyond its on-board sensor range but observed by V2X-equipped connected smart infrastructures or other V2X-enabled connected agents. Figure 1 shows a scenario where a vehicle detects vulnerable road users and share its sensor information with neighboring vehicles. Currently, several ongoing standardization initiatives exist in:

- North America (SAE J3224 [7])
- Europe (ETSI TS 103 324 [6] / TR 103 562 [5])
- China (CSAE 157 [8])

TABLE I: Status of MBD specification per message type

| V2X Message | Misbehavior Detectors Status |
|---|---|
| BSM / CAM | Specified |
| DENM | Specified |
| CPM | Specification is missing |

Even though the standards above have the same purpose, each has its own specification, and thus, might have different cybersecurity threats. In this work, we present our TA of TR 103 562 because it is publicly and freely available.

*2) Security Standard:* ETSI TS 103 759 [9] is a standard under development that defines V2X MBD and reporting activities for CAM and Decentralized Environmental Messages (DENM). The supporting TR 103 460 [10] briefly mentioned the detection and reporting of CPM, but details are out-of-scope of the version 1 of the TS.

### B. Academic work

In [11], Allig et. al. investigated one attacker that forges CPM content. The trustworthiness of CPM is then estimated by assessing data consistency across a pair of entities (using Bayes filter). The simulation shows promising results for detecting this attack and identifying the attacker.

Hadded et al., provided a security analysis of the PAC V2X project [12]. The security analysis covers several V2X applications and V2X messages, including CPM. However, their security analysis of CPM is superficial, considering only two attacks. Our work extended this analysis with 16 attacks that cover different aspects of CPM and CPS.

## III. SYSTEM MODEL

This section provides a system overview, presenting the overall collective perception system used by a Connected Automated Vehicle (CAV), the CPM format, the authentication of CPMs, and some V2X applications consuming CPMs.

### A. Collective Perception System

As seen in Figure 2, sensor data sharing requires a CP system based on V2X communication and local sensors (e.g., camera, RADAR, LiDAR). A CPS relies on transmission and reception of CPMs by CAVs. Before transmitting a CPM, a CAV fuses each sensor's output and generates the CPM containing the fused output. After receiving a CPM, the collective perception system fuses the CPM's data and sensorial objects. This process, named Cooperative Fusion, results in a list of connected and non-connected objects. Connected objects are connected vehicles (CVs) transmitting safety awareness messages (e.g., CAM) to the ego-vehicle (EV). Non-connected objects are objects unable to use V2X communication to send safety awareness messages. Finally, this list of connected and non-connected objects is stored in a Local Dynamic Map (LDM). The latter stores a virtual representation of the EV's surroundings based on received V2X data and EV's sensors (e.g., Camera and GPS). Being able to receive CPMs has two benefits. Firstly, CVs without sensors can perceive non-connected objects. Secondly, CVs with low-cost sensors may have an improved perception thanks to higher-tier sensors equipped on higher-tier CVs.

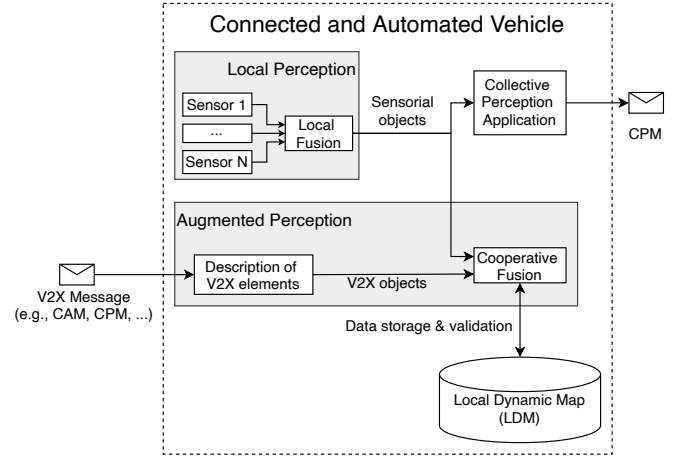

Fig. 2: Collective Perception System [13]

### B. Collective Perception Message Format

Figure 3 shows the structure of a CPM. The greyed fields are optional; white fields are mandatory. CPM consists of an ITS Protocol Data Unit header and containers to include information about the transmitting station, its sensory capabilities, perceived objects and perceived free space.

The *StationDataContainer* and the *ManagementContainer* provide information about the sending station, such as its position and heading. The *SensorInformationContainer* includes details about the onboard sensors of the sending station, such as their identifier, range, and aperture angles. In addition, the *PerceivedObjectContainer* lists relevant objects sensed by the sending station, including their distance from the reporter, speed, dimensions, and other data. Finally, the *FreeSpaceAddendumContainer* lists areas that are unoccupied by an object. It includes the identifier and list of points to denote the free space. A receiver calculates the free space area using simple ray-tracing. In addition, the *FreeSpaceAddendumContainer* can contain the identifiers of sensors linked to the sensor identifiers in *SensorInformationContainer*.

Another important aspect is the transmission rate of CPMs. Since the frequency band dedicated to V2X communication is scarce, it is important to adapt the generation rate and the included objects in the CPMs appropriately to achieve a good trade-off between telematics awareness and channel load. Thereby, preventing channel congestion and a decreased performance of V2X communication. The CPM generation rules currently discussed at ETSI are based on the dynamic properties of the detected objects. If the object's dynamic state has changed in a way that would trigger the generation of a CAM by this object, it is then included in the next CPM [14]. Highly dynamic objects are therefore more often included in transmitted CPMs than slow or static objects.

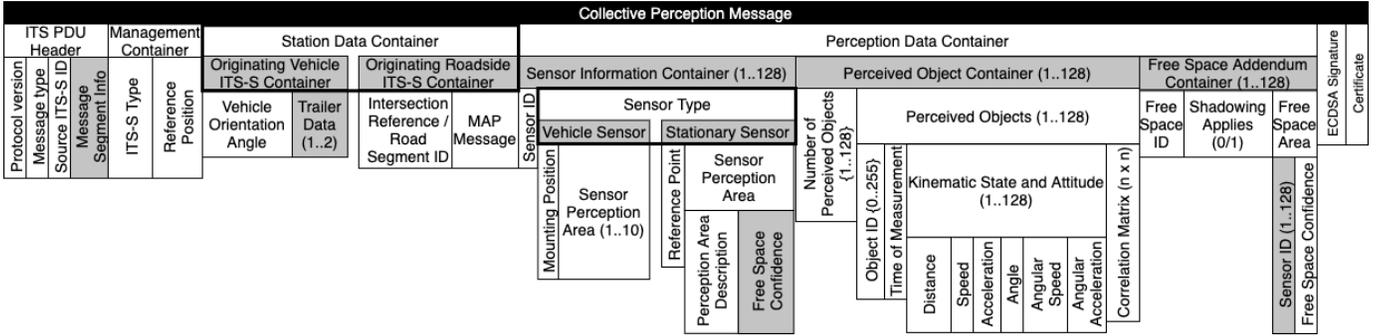

Fig. 3: ETSI Collective Perception Message Format

### C. Authentication of CPMs

The CPS specification includes security requirements such as CPM's integrity, and transmitter's authenticity. Following the IEEE 1609.2 [15], message's integrity and transmitter's authenticity are ensured by digitally signing every CPM sent. Receivers use the transmitter's public key contained in the certificate to verify the digital signature attached to the CPM. This forces the attacker to have valid credentials to perform attacks on CPMs.

### D. V2X Applications

V2X applications rely on V2X messages as an input to warn the driver or to control the vehicle dynamics to avoid road hazard or improving gas consumption. Several safety critical V2X applications would benefit from using CPM [16]:

- Intersection Collision Warning (ICW)
- Emergency Electronic Brake Lights (EEBL)
- Mobile Accessible Pedestrian Signal System (MAPSS)
- Pedestrian in Signalized Crosswalk Warning (PSCW)
- Blind Merge Warning (BMW)

For example, EEBL would benefit from richer information about the location and cause of the event to enhance EV's reaction. From a standard perspective, these cross-application functionalities are unspecified yet.

## IV. ATTACKER MODEL

To facilitate the TA, we formalize the attacker model following the classification proposed in [17].

*Internal versus External:* The internal attacker is an authenticated member of the network that can communicate with other members. The external attacker cannot properly sign her messages, which limits the diversity of attacks. Nevertheless, she can eavesdrop the V2X broadcast communication.

*Malicious versus Rational:* A malicious attacker seeks no personal benefits from the attacks, and aims to harm the members or the functionality of the network. Hence, she may employ any means disregarding corresponding costs and consequences. On the contrary, a rational attacker seeks personal profit and, hence, is more predictable in terms of attack means and attack target.

*Active versus Passive:* An active attacker can generate packets or signals to perform the attack, whereas a passive attacker only eavesdrops the communication channel (i.e., wireless or in-vehicle wired network).

*Local versus Extended:* An attacker can be limited in scope, even if she controls several entities (vehicles or base stations), which make her local. An extended attacker controls several entities that are scattered across the network, thus extending her scope.

*Direct versus Indirect:* A direct attacker reaches its primary target directly, whereas an indirect attacker reaches its primary target through secondary targets. For instance, an indirect attacker may compromise a CPM through a sensor attack.

Figure 4 shows an example of attack on an EEBL application that uses CPM. This example assumes that the dark vehicle fuses its onboard sensors and V2X with equal weight. When detecting conflicting information, it goes in fail-safe mode. As a first step, an attacker (white vehicle) generates multiple ghost vehicles (light gray) at specific locations [17]. Then (step 2), the attacker sends a fake BSM/DENM claiming an emergency brake along with the CPM reporting a stationary (ghost) vehicle ahead. Finally (step 3), the dark gray vehicle detects inconsistencies between its sensor readings and the received information, thus triggering fail-safe mode. This example demonstrates the importance of assessing data trustworthiness and detecting attacks.

## V. THREAT ASSESSMENT

### A. Methodology

Several methodologies assess the risk level for an attack. For example, attack trees were used to formalize attacks on V2V communication [18]. However, in our context, the large number of attacks makes the trees too large and unwieldy. Therefore, our methodology follows a matrix approach based on three criteria: *reproducibility*, *impact*, and *stealthiness* (see Table II). The attack *reproducibility* aims to assess the level of ease to replicate the attack. The *impact* measures how impactful the attack can be on the victim's car and its surrounding vehicles (i.e., criticality and scalability). The attack *stealthiness* assesses the ease by which a driver or a system can detect it. Accordingly, we assess the overall risk level for each threat based on the majority rating among the criteria. For attacks that have all three (High, Medium, Low) ratings in the criteria, the overall rating is taken as Medium.

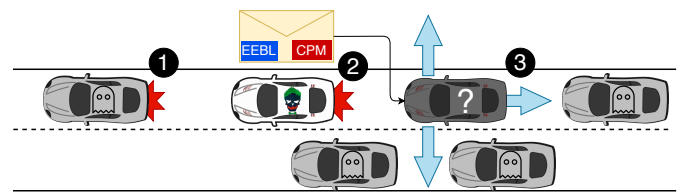

Fig. 4: Attacking EEBL using malicious CPM

TABLE II: Risk ratings and criteria [12]

| Criteria | High | Medium | Low |
|---|---|---|---|
| Reproducibility | The attack is easily reproducible | The attack is reproducible with some limitations | The attack is hard to reproduce due to its complexity or operational cost. |
| Impact | The attack infects the system and can lead to catastrophic damage (e.g., an accident) | The attack infects the system and can lead to moderate damage (e.g., traffic jam) | The attack has no impacts on the system but can inflict minor harm |
| Stealthiness | Unknown attack occurs in certain applications | The attack needs several misbehavior detectors, message types, or data sources to be detected | Broadcasted information readily explain the misbehavior |

### B. Summary

We performed a TA of the ETSI TR 103 562 [5], identifying 16 attacks. Out of the 16 attacks, 13 linked to the TR, and 3 were agnostic to the standards. As a result, we found two high, six medium, and eight low risk attacks.

Although there are more number of medium and low risk attacks, some attacks are very easily reproducible and some have the capability of very high impact to the CPS. We present our analysis in Tables III and IV.

As described in Section IV, the attacker model considered has the ability to modify all of the CPM's containers with any desired value.

One attack on *SensorInformationContainer* considers sensors that can only detect objects until 100 meters but the attacker modifies that value to 200 meters and reports objects at 190 meters. This information is evidently false, but a receiver can't corroborate such information individually.

One attack on *FreeSpaceArea* is an attacker that falsifies a free space where an object is present. The receiving vehicle would only be able to corroborate against this information by coming in line-of-sight (LoS) of the claimed free space.

One attack on *PerceivedObjectContainer* is when an attacker creates fake perceived objects by copying values of other perceived objects (received via CPMs) and modifying only their location information. Another attack on this field is a remote blinding of sensors [19]. In the latter case, the reporting vehicle (i.e., sender of CPM) isn't misbehaving, but is the target of an attack. However, a MBDS could detect that the target should have reported the missing objects, and hence be classified as misbehaving. This example shows the complexity of designing robust MBDS for CPMs.

### C. Conclusion

Most attacks have high *reproducibility* (only one has a medium rating) since they do not require special hardware to perform the attack. The *impact* of 3 out of all attacks in Tables III and IV have high *impact* rating since they have the potential to put the lives of drivers and pedestrians in jeopardy. Lastly, these attacks are lowly rated for *stealthiness* as the attacker would be exposing its certificate in the malicious messages and can be easily detected if the suggested defenses for each attacks are applied.

Although the attacks we developed have high *reproducibility* and *impact*, we have suggested defense mechanisms that should be able to detect such attacks and help report the malicious actors. These defense mechanisms mainly require redundant information from other honest actors surrounding the target vehicle or redundant sensors on the target vehicle. However, as discussed in Section VI, the functional standards are focusing on redundancy mitigation techniques to reduce channel load. Thus, the defense mechanisms can only be practically applied if the standards allow room for redundant information.

## VI. DISCUSSION

In this section, we propose standard-related directions to address some of the security gaps identified by the TA.

### A. Misbehavior Detectors and Reporting

ETSI TR 103 460 and TS 103 759 list a set of misbehavior detectors for CAM. Currently, the TS draft does not specify detectors for the CPM, leaving that for a future version. However, we can assume that detectors (designed for CAM) will be applicable to CPM too. For instance, in TR 103 460, the detector, named *implausible speed*, will be the same for both CPM and CAM.

Additional detectors specific to CPM will be needed though. A potential detector could use *SensorInformationContainer* to detect fake perceived objects. Indeed, an attacker can generate randomly positioned perceived objects in the *PerceivedObjectContainer*. A detector should verify if each perceived object is within the sensory perception area. A perceived object outside the sensor perception area should not have been detected by the sensor, and thus, most likely does not exist. In a similar fashion, a CPM detector could verify if two CPMs from different senders are consistent. For instance, a perceived object within the perceived area of vehicle A and vehicle B should be part of the CPM sent by vehicle B. This observation could mean vehicle A have inserted a fake perceived object or vehicle B suppressed the perceived object. Thus, an absence of consistency between the two CPMs may increase at least the suspicious level for both reporting vehicles.

After being detected, a misbehavior report (MBR) may be generated and sent to authorities for further investigation. The ASN.1 definition specified in TS 103 759 should be flexible enough to allow for CPM detectors.

### B. Tension between redundancy mitigation and MBD

If multiple stations perceive the same (physical) object, redundant and unnecessary frequent updates about that object will be broadcast, thereby increasing the network channel load. To address this issue, ETSI CPS defined redundancy mitigation rules. These can be frequency-based, dynamics-based, or confidence-based, and triggered when the observed channel busy ratio is higher than a predefined threshold [5]. However, as noted earlier, the redundancy can be useful to detect misbehaviors. An interesting work item could be to study this trade-off, and define an approach to balance between redundancy and channel congestion.

### C. Use of CPM as data source for V2X MBD (and vice versa)

It can be tempting to use CPM as data source to detect malicious CAM (or to use CAM to detect malicious CPM). For instance, a perceived and connected object in a CPM may have sent CAM information that are consistent with the corresponding CPM. However, the use of other message is not trivial because the CAM and the CPM are received at different moment in time. A motion prediction algorithm (e.g., Kalman Filter) could tackle this issue. However, the standard should make clear if all vehicle shall use the same prediction algorithm, and shall provide the temporally synchronized BSM and CPM in the corresponding MBR. The specification of this approach might impact the ASN.1 definition of the MBR.

To further improve the CPMs' trustworthiness and prevent attacks on *SensorInformationContainer* is could be useful to extend the IEEE 1609.2 certificate format to include EV's capabilities. This would allow for (authenticated) attestation of sensing capabilities.

### D. Misbehavior Detection for sensors and fusion

The V2X module of a CAV assumes trustworthy sensor data. This assumption is strong as attacks on automotive RADAR, LiDAR, and camera have been demonstrated [19]. As highlighted in Tables III and IV, a MBDS using local sensors cannot ensure the plausibility of a CPM content. Indeed, if sensors can be fooled or jammed by an external attacker, then sensors cannot be a reliable data source for a MBDS. Standardizing misbehavior detectors for sensor will allow a transmitter to insert trusted sensor data in a CPM before its transmission. For instance, a machine learning module could verify if the object detected by a sensor has a plausible location and motion [20]. Such standardization effort could happen in the ISO TC22 SC32 committee as part of the future ISO 5083.

## VII. Conclusion

CPS offers to V2X-equipped vehicles the ability to exchange richer data to improve further their telematics awareness and safety. However, the security of CPM is mandatory to guarantee quality data. Standardization efforts of CPS and V2X MBD (separately) are ongoing worldwide, but misbehavior protection in CPS still has to be addressed. In this paper, we provided a summary of a TA done on ETSI TR 103 562, which identified 16 attacks with mainly low to medium risk level. From this assessment, we proposed four work items for consideration in ongoing standardization efforts. We hope this work could serve as a starting point to tackle the question of CPS security by standard organizations and regulators.

TABLE III: Threat analysis of use-cases from ETSI TS 103 324

| Ref. | Use Case | Attacks | Defense | Risk |
|---|---|---|---|---|
| A | ObjectID is assigned to each detected remote V2X object by the transmitter | Attacker generates enough objects to exceed the number of trackable objects | only track relevant objects. Use noise cancellation techniques to filter out transient attacks (similar to RADAR processing). | Low. **Reproducibility:** Medium. Generating large number of objects and transmitting them over multiple CPMs will require high processing power. **Impact:** Low. Other remote V2X objects may report the objects missed by ego V2X object. **Stealthiness:** Low. Attacker is detectable through its certificate in CPM. |
| B | | Attacker spoofs objects such that transmitter associates the spoofed object with previously transmitted ObjectID | Objects don't change in size, if that is looked for in change, this attack may be detected | Medium. **Reproducibility:** Medium. The attacker needs to be the first one to transmit a CPM with the ObjectID of interest. This is not always possible due to inherent congestion control mechanisms in wireless networks. **Impact:** Medium. Although the ObjectID of interest is spoofed, the real object may be assigned a new one. This may cause that object to be reported by others as malicious, raising many false positives in the revocation system. **Stealthiness:** Low. All receivers will know who transmitted the false information since CPM is broadcasted to all receivers in range. |

TABLE III: Threat analysis of use-cases from ETSI TS 103 324 (continued)

| Ref. | Use Case | Attacks | Defense | Risk |
|------|----------|---------|---------|------|
| C | Receiver should listen for 1 second (at least) to receive all objects in PerceivedObjectContainer | Attacker starts transmission between this 1 second. | Detect such energy blast as spectrum misbehavior | Medium. **Reproducibility:** Medium. The attacker would need special hardware (such as a software defined radio) to be able to circumvent congestion control mechanisms of the lower layers in wireless communication. **Impact:** Medium. The receiver will only be able to receive partial information but no malicious information since the attacker is only disrupting the connection between the two original communicating stations. **Stealthiness:** Low. The attacker can be identified through the certificates in its messages. If the attacker is transmitting bogus information, triangulation techniques on the transmitter's energy source may be used to pin-point the attacker. |
| D | Objects are classified as Person/Animal or other | Put up mannequins to be detected as person/animal | None needed | Low. **Reproducibility:** Low. Attacker needs to bring mannequins on the road. **Impact:** Low. Mannequins may still be detected as objects. Even if they are detected as a person, vehicles will know their presence and act accordingly. **Stealthiness:** Low. Attacker would have to be physically present to put and move mannequins. |

TABLE III: Threat analysis of use-cases from ETSI TS 103 324 (continued)

| Ref. | Use Case | Attacks | Defense | Risk |
|---|---|---|---|---|
| E | If classified object is not a person or animal estimated orientation change > 4 degrees | Attacker uses obscure painting techniques to confuse the camera about its heading | Rely more on lidar and infer heading by analyzing movement | Medium. **Reproducibility:** Low. Attacker needs to know the painting techniques that may fool camera, lidar and radar systems altogether. **Impact:** Medium. Camera may get fooled due to optical illusions created by the paintings but lidar and radar will not have any effect on their performance. **Stealthiness:** Low. The obscurely painted vehicle will be physically recognizable. |
| F | If classified object is a person or animal seen 1st time | Attacker puts a moving mannequin/scarecrow on wheels to move as fast as a small scooter. | Such an "object" will still be reported in a CPM so it should not affect the system so much. | Low. **Reproducibility:** Low. Attacker needs to bring mannequins on the road. **Impact:** Low. Mannequins may still be detected as objects. Even if they are detected as a person, vehicles will know their presence and act accordingly. **Stealthiness:** Low. Attacker would have to be physically present to put and move mannequins. |
| G | If classified object is a person or animal if even one person/animal was not in CPM since 500ms, all person and animal object have to be included in currently generated CPM | Attacker jumps in and out of LoS to make vehicle re-transmit everyone, using up resources | None needed | Low. **Reproducibility:** Low. Attacker needs to be physically present and perform movements at the right time to affect enough number of transmitting V2X objects. **Impact:** Low. Attacker's movement may still be missed if there are many people around the attacker. **Stealthiness:** Low. Attacker will be physically present and recognizable. |

TABLE III: Threat analysis of use-cases from ETSI TS 103
324 (continued)

| Ref. | Use Case | Attacks | Defense | Risk |
|---|---|---|---|---|
| H | *Sensor Information Container* can contain sensor capability of receiving V2X object. | Report object at 100m whereas it is at 10m or 200m | Correlate with multiple vehicle's information | High. **Reproducibility:** High. A man-in-the-middle attacker can encode different values than reality into the Sensor-InformationContainer before signing and transmitting. **Impact:** High. If an object is spoofed to be suddenly in front of another vehicle. It may cause sudden reactions by the vehicle system, causing collisions. **Stealthiness:** Medium. A quick check with on-board sensors should reveal anomaly in CPM information. |
| I | | Report 1m capability and report object at 5000m | None needed | Low. **Reproducibility:** High. An attacker with man-in-the-middle capability should be able to modify message contents for this attack. **Impact:** Low. A simple check against capability of the sensor should reveal the inconsistency. **Stealthiness:** Low. Attacker will have to transmit this in its own CPM and hence will be recognizable through its certificate. |

TABLE III: Threat analysis of use-cases from ETSI TS 103 324 (continued)

| Ref. | Use Case | Attacks | Defense | Risk |
|------|----------|---------|---------|------|
| K | ConfirmedFreeSpace calculation MAY be included and calculated using ray tracing and shadowing | Send wrong Confirmed-FreeSpace to make receiver believe of free space | None needed | Medium. **Reproducibility:** High. An attacker with man-in-the-middle capability will be able to modify message contents for this attack. **Impact:** Medium. If a transmitting V2X object wants to make a maneuver based on the free space, it will not be able to do so. Hence, the attacker would be successful in hindering the transmitting V2X objects' operation. **Stealthiness:** Medium. If the transmitting V2X object eventually comes to a position where it is able to see its area of interest directly, it will be able to detect the attack and recognize the attacker through its certificate. |

TABLE III: Threat analysis of use-cases from ETSI TS 103 324 (continued)

| Ref. | Use Case | Attacks | Defense | Risk |
|------|----------|---------|---------|------|
| L | | Blinding lidar and camera using lasers | Image processing to detect camera blinding. For LIDAR, we need encoded pulse. Out of scope for AMPS-V2X. | Medium.<br>**Reproducibility:** Low. Modifying radar or lidar pulses requires sophisticated hardware that understands the signal quickly to modify them meaningfully. Cost of the hardware may also be a concern.<br>**Impact:** High. LIDAR is a highly relied upon system for object detection due to its precision. RADAR is highly relied upon by ACC systems due to its sensitivity to change in distance. Hence, the impact of blinding these systems may have catastrophic effects on the whole stack of an autonomous driving system.<br>**Stealthiness:** Medium. These attacks do not require the attacker to reveal its identity, making the attacker extremely hard to find. Although the attack itself is detectable as a sort of DoS on LIDAR and RADAR. |
| M | CPM should be generated within 50ms (generation time = CPM generation trigger - CPM handoff to Network & Transport layer) | Attacker performs DoS of CPM | Observe too many CPM (or too few) | Low.<br>**Reproducibility:** Low. Attacker needs specialized hardware with enough computation power to generate substantially large CPMs quickly.<br>**Impact:** Medium. The channel may get congested due to this attack starving communication resources for CPS.<br>**Stealthiness:** Low. Transmitting V2X objects will be able to recognize the attacker through the certificates in its CPMs. |

TABLE III: Threat analysis of use-cases from ETSI TS 103 324 (continued)

| Ref. | Use Case | Attacks | Defense | Risk |
|------|----------|---------|---------|------|
| N | | Attacker blasts energy at the time when a new CPM is about be generated to interfere with reference position determination using GPS | Report such an abnormal energy blast for physical inspection | Medium **Reproducibility:** Medium. An attacker with high energy producing device can perform this kind of attack. However, the attacker would need to synchronize the attack with CPM generation by the target, which might be difficult to achieve. **Impact:** High. All surrounding vehicles may be affected by such energy bursts, bringing the communication infrastructure for CPS to a halt. **Stealthiness:** Low. High energy bursts can be easily detected by receivers as they already do so for any message they receive. |

TABLE IV: Threat analysis of use-cases agnostic to standards

| Ref. | Use Case | Attacks | Defense | Risk |
|------|----------|---------|---------|------|
| A | RSU will not transmit CPM for vehicles that they have received a CAM from | Transmit CAMs for other objects to make them disappear from CPM (falsify telematic information to match that of the other object) | The vehicle is already aware of all the surrounding objects | Low. The impact of this attack is very low. **Reproducibility:** Medium. Attacker will have to use its own certificates for CAMs of transmitting V2X object. **Impact:** Low. Other vehicles will still know the presence of transmitting V2X object, hence this attack will only save the network bandwidth and not attack it. **Stealthiness:** Medium. Receiving V2X objects may not be able to infer location of attacker but global V2X systems like a misbehavior authority may be able to do so by looking at any reported certificates in misbehavior reports. |

TABLE IV: Threat analysis of use-cases agnostic to standards
(continued)

| Ref. | Use Case | Attacks | Defense | Risk |
|---|---|---|---|---|
| B | Objects are reported with a time offset based on synchronization with a time clock | Attacker can target time synchronization attack to offset generation time so that the receiver thinks a reported object is farther/closer than real position or spoof speed etc. | Make sure the connection to clock sync servers is secured | Low. **Reproducibility:** Low. Accessing a secure time server will require high-level physical access. These time servers are usually well protected against web-based attacks as well. **Impact:** High. Loss of time synchronization will cause most applications and time-distance based calculations to work falsely. **Stealthiness:** Low. Only a security analysis of the time server may uncover any malicious activity. |
| C | | Moving box with the size of a car/truck on wheels | No real threat with this attack since a truck-sized box is also physically detectable | Low. **Reproducibility:** Low. Attacker needs a lot of equipment to build huge boxes to represent a truck. **Impact:** Low. The box will still be detected as an object and hence vehicles will be able to maneuver accordingly. **Stealthiness:** Medium. If a box is able to fool transmitting V2X object as being a truck, it is still a relatively stealthy attack. |